\documentclass[epj]{webofc}
\usepackage[varg]{txfonts}   
\usepackage{subfigure}
\usepackage{amsmath}
\usepackage{lineno}
\begin{document}
%
\selectlanguage{english}
\title{Prospects for Higgs CP property measurements at the LHC}
%
%

\author{Xin Chen\inst{1,2,3}{\fnsep\thanks{\email{xin.chen@cern.ch}}, On behalf of the ATLAS and CMS Collaborations}
}

\institute{Department of Physics, Tsinghua University, Beijing 100084, China
\and Collaborative Innovation Center of Quantum Matter, Beijing 100084, China
\and Center for High Energy Physics, Peking University, Beijing 100084, China
}

\abstract{%
This document is prepared for the LCWS2016 conference proceedings. It reviews the current results on the Higgs CP property measurements from both ATLAS and CMS experiments in the Higgs to diboson decays, and in the Vector Boson Fusion production of Higgs via the ditau decay channel. The projected sensitivity of the ditau signal in HL-LHC is briefly discussed. Finally, it gives the prospects for the CP measurement in the Higgs to ditau decay with the tau substructure developments at both collaborations.
}
\maketitle
\section{Introduction}
\label{intro}
After the Higgs was discovered in 2012 \cite{higgs1,higgs2}, understanding its properties, and looking for any possible deviations of its properties from the SM prediction, becomes a very important task of LHC. This includes measuring the its spin (scalar or tensor), its mass which has important implications for the vacuum stability and cosmology, its flavor couplings to look for Beyond Standard Model (BSM) signatures of flavor changing effect, its exotic decay modes such as $H\to\text{invisible}$, and its CP property which is related to the matter-antimatter imbalance in the universe. If the Higgs is in a pure CP eigen state, we will find if it is CP even or odd. On the other hand, if it is in a CP mixture, we will measure the CP mixing angle. This is a more exciting scenario, as it gives rise to an extra CP violation source, and the current known source (a single complex phase in CKM) is too small to explain the matter-antimatter imbalance.

\section{CP test in the bosonic decays of the Higgs}
\label{boson_decay}

\begin{figure}[h]
\centering
\includegraphics[width=6.cm,clip]{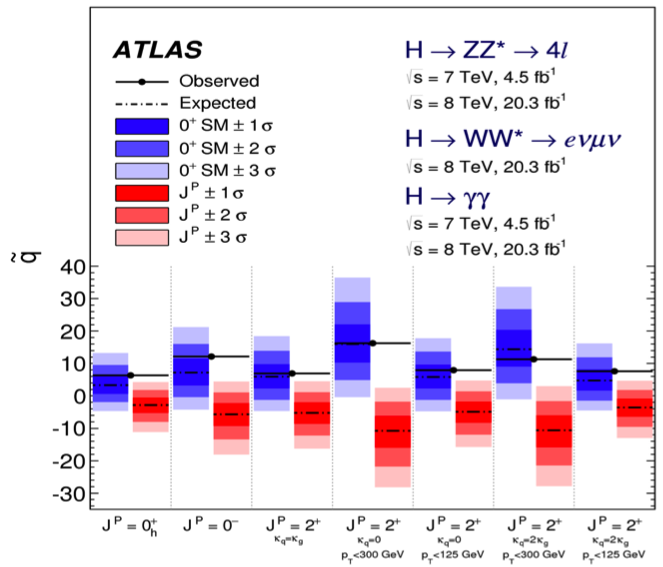}
\caption{ The test results of Higgs with different spin and CP states at ATLAS \cite{EFT1}. The measurements are indicated by the black solid bars. }
\label{fig:fig1}
\end{figure}

\begin{figure}[h]
\centering
\includegraphics[width=12.cm,clip]{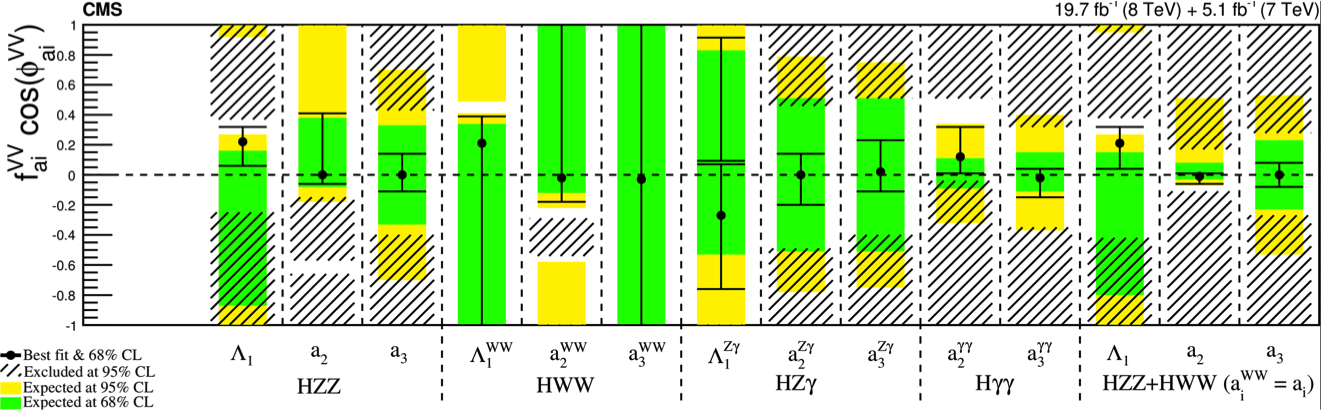}
\caption{ The bounds on the coefficients of the different BSM terms for $H\to ZZ^*$, $H\to WW^*$, $H\to Z\gamma$ and $H\to \gamma\gamma$ decays at CMS \cite{EFT2}. }
\label{fig:fig2}
\end{figure}

The CP symmetry of the Higgs coupling to vector bosons are parametrized with an effective Lagrangian by the ATLAS collaboration as \cite{EFT1,EFT1b}
\begin{eqnarray}
\begin{array}{ll}
\mathcal{L}_0^V = & \left\{ c_\alpha \kappa_{SM}  \left[ \frac{1}{2} g_{HZZ} Z_\mu Z^\mu +  g_{HWW}W_\mu^+ W^{-\mu} \right] -\frac{1}{4} \frac{1}{\Lambda}  \left[ c_\alpha \kappa_{HZZ} Z_{\mu\nu} Z^{\mu\nu} + s_\alpha \kappa_{AZZ} Z_{\mu\nu} \tilde{Z}^{\mu\nu}  \right] \right.  \\
& \left. -\frac{1}{2} \frac{1}{\Lambda} \left[ c_\alpha \kappa_{HWW}  W_{\mu\nu}^+ W^{-\mu\nu} + s_\alpha \kappa_{AWW}  W^+_{\mu\nu} W^{-\mu\nu} \right] \right\} X_0 ,
\end{array}
\label{eq:eq1}
\end{eqnarray}
where $\alpha$ is the CP mixing angle, $c_\alpha = \cos{\alpha}$, $s_\alpha = \sin{\alpha}$, $\Lambda$ is scale cut-off for dim-6 operators, $\tilde{V}_{\mu\nu}(V=W/Z)$ is the dual tensor of $V_{\mu\nu}$, and $X_0$ is the neutral resonance such as the Higgs. The terms with $1/\Lambda$ coefficients are BSM dim-6 operators, with the $c_\alpha$ ($s_\alpha$) terms representing the BSM CP even (odd) contributions. Based on Eq.~\ref{eq:eq1}, the pure CP states of Higgs can be parametrized as in Tab.~\ref{tab:tab1}. CMS collaboration used a similar expression, and the part with $Z$ boson terms only are \cite{EFT2}
\begin{equation}
\mathcal{L}_{HVV} \sim a_1 \frac{m_Z^2}{2} HZ^\mu Z_\mu - \frac{\kappa_1}{\Lambda_1^2} m_Z^2 H Z_\mu \square Z^\mu - \frac{1}{2} a_2 H Z^{\mu\nu}Z_{\mu\nu} - \frac{1}{2} a_3 HZ^{\mu\nu}\tilde{Z}_{\mu\nu}, 
\label{eq:eq2}
\end{equation}
\begin{table}[h]
\caption{ The definitions of $HVV$ coupling coefficients for pure Higgs CP states (ATLAS) \cite{EFT1}. }
\label{tab:tab1}
\centering 
\begin{tabular}{cccccc}
\hline \hline
$J_P$ & Model & \multicolumn{4}{c}{Choice of tensor couplings} \\
& & $\kappa_{SM}$ &  $\kappa_{HVV}$ &  $\kappa_{AVV}$ & $\alpha$ \\ \hline
$0^+$ & Standard Model Higgs boson & 1 & 0 & 0 & 0 \\
$0_h^+$ & BSM spin-0 CP-even & 0 & 1 & 0 & 0 \\
$0^-$ & BSM spin-0 CP-odd & 0 & 0 & 1 & $\pi/2$ \\ \hline \hline
\end{tabular}
\end{table}
Both ATLAS and CMS have tested the CP states of the Higgs assuming it is in a pure CP state. Both experiments used a Matrix Element (ME) method to test the $0^+$ state against $0^-$.  ATLAS combined three channels in the test: $H\to WW\to e\nu \mu\nu$ 0-jet category, $H\to\gamma\gamma$ and $H\to ZZ\to 4l$. The results are as shown in Fig.~\ref{fig:fig1}. CMS combined $H\to WW, ZZ, Z\gamma, \gamma\gamma$ channels and for the pure CP test \cite{EFT2}.  The CP odd $0^-$ model is excluded at CL better than $99.9\%$ by both experiments.
When the Higgs is in a CP-mixed state, both experiments have made constraints on the coefficients of different couplings in Eq.~\ref{eq:eq1} and \ref{eq:eq2}, as shown in Tab.~\ref{tab:tab2} and Fig.~\ref{fig:fig2}. One can conclude that the non-SM tensor couplings are consistent with zero for both experiments.
\begin{table}[h]
\caption{ The best-fit and excluded ranges for the coefficients of the BSM CP-even and CP-odd terms in Eq.~\ref{eq:eq1} at ATLAS, with $H\to ZZ^*$ and $H\to WW^*$ combined \cite{EFT1}.}
\label{tab:tab2}
\centering 
\begin{tabular}{ccccc}
\hline \hline
Coupling ratio & \multicolumn{2}{c}{Best fit value} & \multicolumn{2}{c}{$95\%$ CL Exclusion Regions} \\
Combined & Expected & Observed & Expected & Observed \\ \hline
$\tilde{\kappa}_{HVV}/\kappa_{SM}$ & 0.0 & -0.48 & $(-\infty,-0.55]\cup[4.80,\infty)$ & $(-\infty,-0.73]\cup[0.63,\infty)$ \\
$(\tilde{\kappa}_{AVV}/\kappa_{SM})\cdot\tan\alpha$ & 0.0 & -0.68 & $(-\infty,-2.33]\cup[2.30,\infty)$ & $(-\infty,-2.18]\cup[0.83,\infty)$ \\
\hline \hline
\end{tabular}
\end{table}

Another important way to search for Higgs CP violation is in the Vector-Boson-Fusion (VBF) production of the Higgs. One of the most promising channel for this search is $H\to\tau\tau$ decay. The combined Run-1 sensitivity of ATLAS and CMS has already exceeded $5\sigma$ \cite{Htt}. This channel is special in that it is sensitive to CP in both $HVV$ and $Hf\bar{f}$ couplings. In the 2HDM model, the rate of $H\to\tau\tau$ is often enhanced, and mixed CP for Higgs can be accommodated.  

Omitting the BSM CP-even terms, the $HVV$ coupling can be written as \cite{OO}
\begin{equation}
\mathcal{L}_{eff} = \mathcal{L}_{SM} + \tilde{g}_{HAA}H\tilde{A}_{\mu\nu}A^{\mu\nu} + \tilde{g}_{HAZ}H\tilde{A}_{\mu\nu}Z^{\mu\nu} + \tilde{g}_{HZZ}H\tilde{Z}_{\mu\nu}Z^{\mu\nu} + \tilde{g}_{HWW}H\tilde{W}^+_{\mu\nu}W^{-\mu\nu}.
\label{eq:eq3}
\end{equation}
Since the vector bosons in the process are not directly observable, it is easier to treat the CP-odd terms as one by the assumption in Eq.~\ref{eq:eq4} about the relations of the coefficients:
\begin{equation}
\tilde{g}_{HAA} =  \tilde{g}_{HZZ} = \frac{1}{2} \tilde{g}_{HWW} = \frac{g}{2m_W} \tilde{d}, ~~\text{and}~~\tilde{g}_{HAZ}=0.
\label{eq:eq4}
\end{equation}
Traditionally, the signed $\Delta\phi$ between the two tagging in the VBF process are used as the CP sensitive variable \cite{VBF}. ATLAS used a different variable, the Optimal Observable (OO), which is expected to perform better. It is based on the ME of
\begin{equation}
\mathcal{M} = \mathcal{M}_{SM} + \tilde{d} \cdot \mathcal{M} _{CP-odd},  ~~\text{and}~~
\left|\mathcal{M}\right|^2 = \left|\mathcal{M}_{SM}\right|^2 + \tilde{d} \cdot 2 Re\left( \mathcal{M}_{SM}^*\mathcal{M}_{CP-odd} \right) +  \tilde{d}^2 \left| \mathcal{M}_{CP-odd} \right|^2,
\label{eq:eq5}
\end{equation}
and the OO is defined as
\begin{equation}
OO = \frac{2 Re \left(\mathcal{M}_{SM}^* \mathcal{M}_{CP-odd}  \right)}{\left| \mathcal{M}_{SM} \right|^2}.
\label{eq:eq5b}
\end{equation}

\begin{figure}[h]
\centering
\includegraphics[width=6.cm,clip]{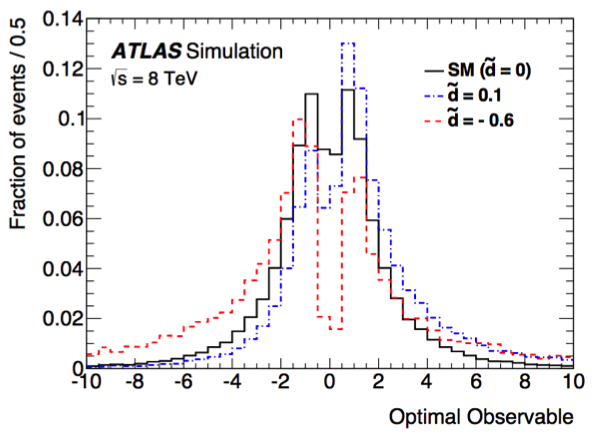}
\caption{ The distributions of OO for the SM ($\tilde{d}=0$) and two other values of $\tilde{d}$ in the VBF $H\to\tau\tau$ signal at ATLAS \cite{OO}. }
\label{fig:fig3}
\end{figure}

\begin{figure}[h]
\centering
\includegraphics[width=5.7cm,clip]{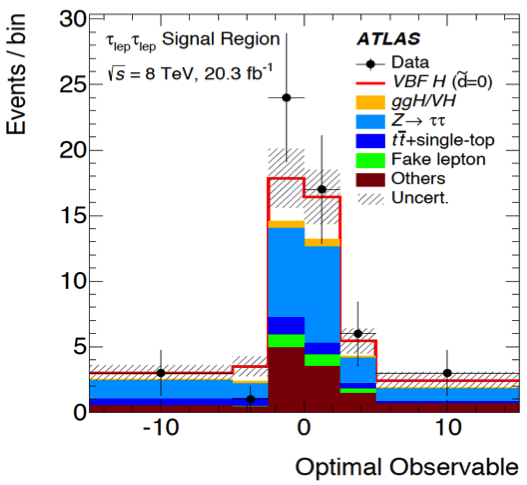}\hspace{5mm}
\includegraphics[width=6cm,clip]{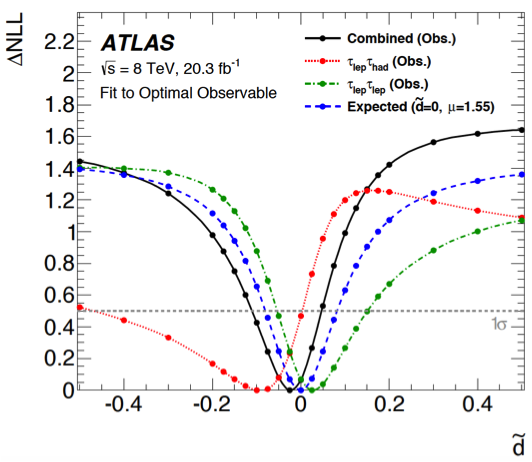}
\caption{ The distributions of the OO in the $H\to\tau\tau\to ll+4\nu$ subchannel (left), and the $\Delta$NLL for different $\tilde{d}$ hypotheses (right) at ATLAS \cite{OO}.}
\label{fig:fig4}
\end{figure}

\begin{figure}[h]
\centering
\includegraphics[width=5.0cm,clip]{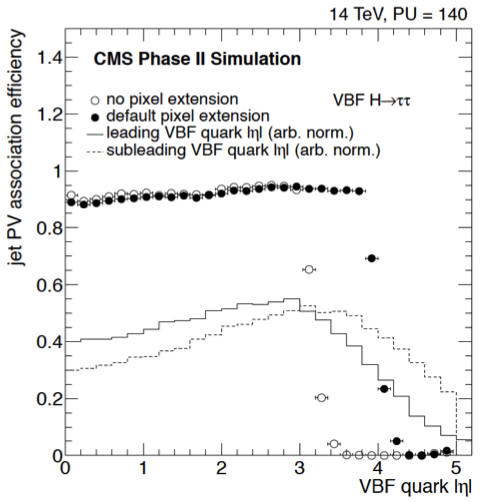}
\caption{ The jet-PV association efficiency as a function of the VBF quark $\eta$ in the $H\to\tau\tau$ channel at CMS. The improvement is evident with the tracking volume extension \cite{HL2}. }
\label{fig:fig5}
\end{figure}

With all 4-momenta of the final state particles (the Higgs and two tagging jets) measured, the LO ME of SM and CP-odd can be calculated from HAWK \cite{hawk}, and then the OO can be calculated per event. As Fig.~\ref{fig:fig3} shows, of there is no CP violation, the mean of the OO distribution should be zero. For positive (negative) CP-odd component (determined by $\tilde{d}$), its mean will be shifted to positive (negative) values. This method can be applied to other decays such as $H\to\gamma\gamma$. 
To obtain a pure signal sample, a cut is first made on the Multi-Variate-Analysis (MVA) output score. A signal to background ratio of about 0.3 is achieved by this cut. Next, a likelihood fit to the OO distribution is performed to find the best value for $\tilde{d}$. Figure~\ref{fig:fig4} shows the OO distribution in the $H\to\tau\tau\to ll+4\nu$ subchannel, and the increase of the negative-log-likelihood $\Delta$NLL with respect to its minimum in the likelihood scan.  Each point in the plot indicates a $\Delta$NLL calculated with a particular hypothesis template and the data. The $68\%$ CL interval is found by the intersection points of the $\Delta$NLL curve and the horizontal line at $\Delta\text{NLL}=0.5$. The result is $-0.11<\tilde{d}<0.05$. While the result shows that the CP-odd contribution is consistent with zero, it is about 10 times tighter than the one shown in Tab.~\ref{tab:tab2} for the $HVV$ coupling. 
In the HL-LHC phase, the Missing Transverse Energy and the di-tau mass resolution will degrade as the pileup increases. The signal-background discrimination increases with the tracking coverage, as the VBF forward jets are better separated from the pileup jets with the association between the jets and the primary vertex, shown in Fig.~\ref{fig:fig5}. Assuming zero theory uncertainty, the uncertainty on the signal strength $\mu$ at 8-18\% (2-5\%) can be achieved with the ATLAS $\tau_l \tau_h$ alone \cite{HL1} (CMS $\tau\tau$ inclusive \cite{HL2}). The projected uncertainty of $\mu$ for different tracking extensions at ATLAS is listed in Tab.~\ref{tab:tab3}. Thus, we would expect a corresponding increase in the precision for the Higgs CP measurement in the VBF channel due to both increased data and extended tracking coverage.
\begin{table}[h]
\caption{ The projected uncertainty on the signal strength for different scenarios of tracking coverage and forward pileup jet rejection at HL-LHC for ATLAS \cite{HL1}.}
\label{tab:tab3}
\centering 
\begin{tabular}{c|ccc}
\hline \hline
forward pile-up jet rejection & $50\%$ & $75\%$ & $90\%$ \\ \hline \hline
forward tracker coverage & \multicolumn{3}{c}{$\Delta\mu$} \\ \hline 
Run-I tracking volume & \multicolumn{3}{c}{0.24} \\ 
$\eta<3.0$ & \multicolumn{1}{c| }{0.18} & \multicolumn{1}{c| }{0.15} & \multicolumn{1}{c}{0.14} \\
$\eta<3.5$ & \multicolumn{1}{c| }{0.18} & \multicolumn{1}{c| }{0.13} & \multicolumn{1}{c}{0.11} \\
$\eta<4.0$ & \multicolumn{1}{c| }{0.16} & \multicolumn{1}{c| }{0.12} & \multicolumn{1}{c}{0.08} \\
\hline \hline
\end{tabular}
\end{table}

\section{CP test in the fermionic decays of the Higgs}
\label{fermion_decay}

The CP-odd Yukawa coupling can enter the Lagrangian as a dimension-4 operator as in Eq.~\ref{eq:eq6}, thus the $Hf\bar{f}$, especially the  $H\tau\tau$ coupling, is a sensitive probe of CP at tree-level rather than the loop level as with the dimension-6 operators in the $HVV$ coupling.
\begin{equation}
\mathcal{L} = -g_\tau \left( \cos\phi\bar{\tau}\tau + i\sin\phi\bar{\tau}\gamma_5\tau \right)h,
\label{eq:eq6}
\end{equation}
where $\phi$ is the mixing angle between the CP even and odd terms. The CP of $H\tau\tau$ coupling can be distinguished by the transverse tau spin correlations, as the decay width is proportional to
\begin{equation}
\Gamma(H,A\to\tau^-\tau^+) \sim 1 - s_z^{\tau-}s_z^{\tau+} \pm s_T^{\tau-}s_T^{\tau+} ,
\label{eq:eq7}
\end{equation}
where $s_{z,T}$ are the tau spin components in the longitudinal and transverse directions with respect to the tau momentum. As a result, the angle between the planes spanned by the tau and its charged track is sensitive to the CP. For example, in the $\tau\to\pi\nu$ decay, one can look at the angle between tau decay planes to extract the CP mixing angle $\phi$:
\begin{equation}
\frac{d\Gamma\left(H\to\tau\tau\to\pi^+\pi^-+2\nu \right)}{d\phi_{CP}} \sim 1-\frac{\pi^2}{16}\cos\left(\phi_{CP}-2\phi \right),
\label{eq:eq8}
\end{equation}
where $\phi_{CP}$ is the angle between the tau decay planes in the ditau rest frame.
%
%

\begin{figure}[h]
\centering
\includegraphics[width=4.7cm,clip]{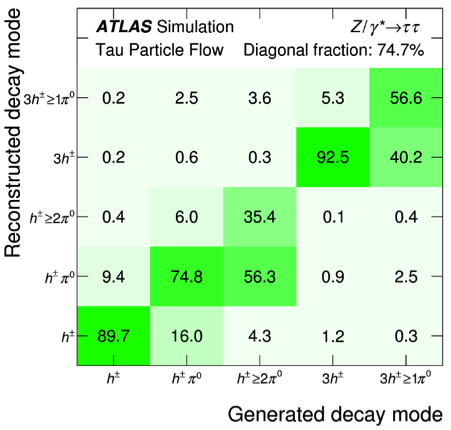}\hspace{5mm}
\includegraphics[width=4.7cm,clip]{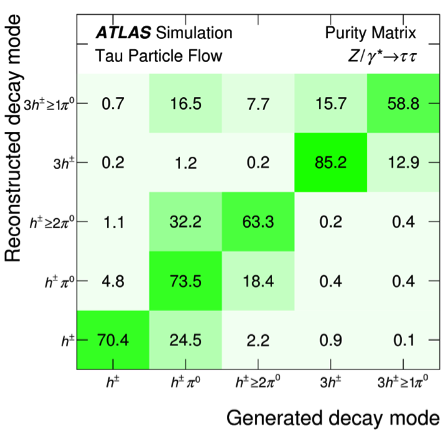}
\caption{ The efficiencies for the generated tau decay modes to be reconstructed as different measured modes (left), and the fractions of different reconstructed modes coming from different generated modes indicating the purity of the reconstruction (right), at ATLAS \cite{sub1}. }
\label{fig:fig7}
\end{figure}

\begin{figure}[h]
\centering
\includegraphics[width=4cm,clip]{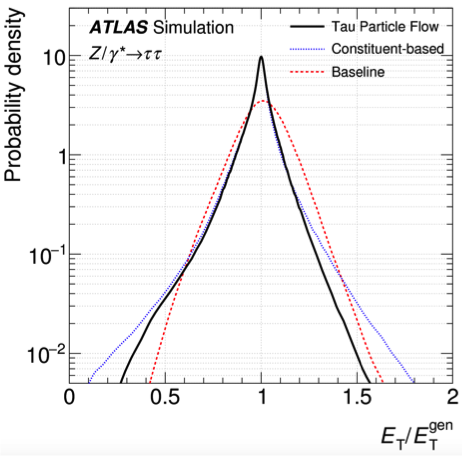}\hspace{3mm}
\includegraphics[width=4cm,clip]{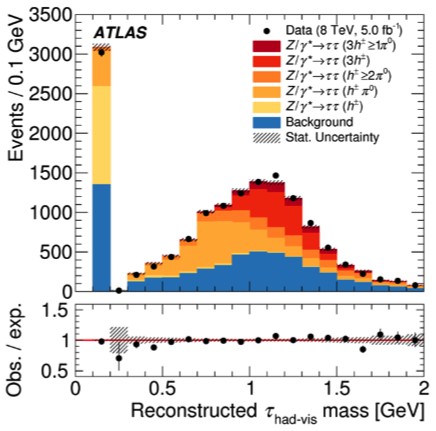}\hspace{3mm}
\includegraphics[width=4cm,clip]{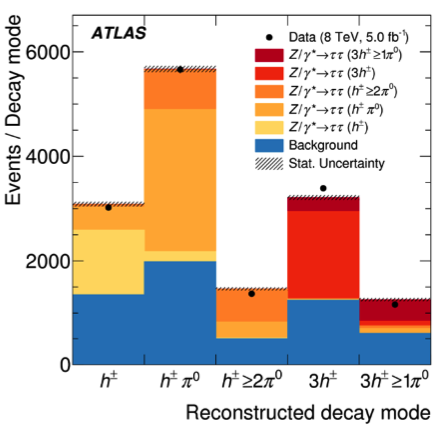}
\caption{ The resolution of the reconstructed tau energy (left), the mass (middle) and modes (right) of the reconstructed taus with substructure at ATLAS \cite{sub1}. }
\label{fig:fig8}
\end{figure}

\begin{figure}[h]
\centering
\includegraphics[width=4cm,clip]{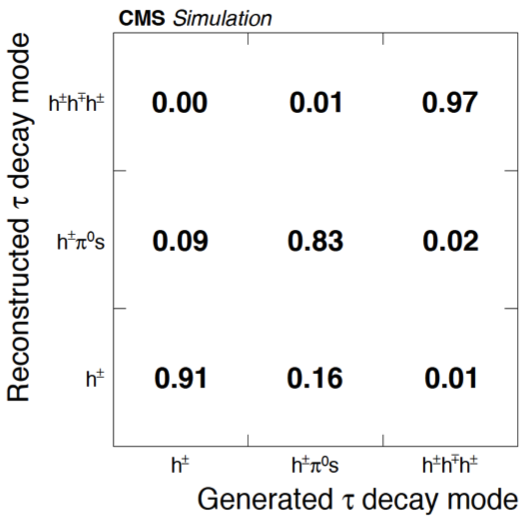}\hspace{3mm}
\includegraphics[width=4.0cm,clip]{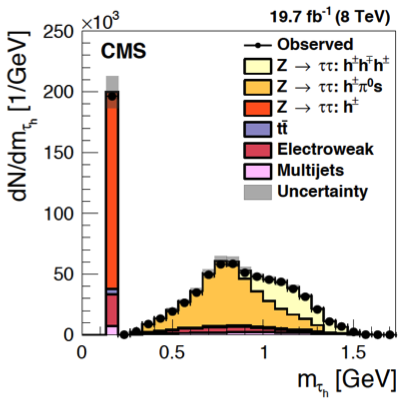}\hspace{3mm}
\includegraphics[width=3.8cm,clip]{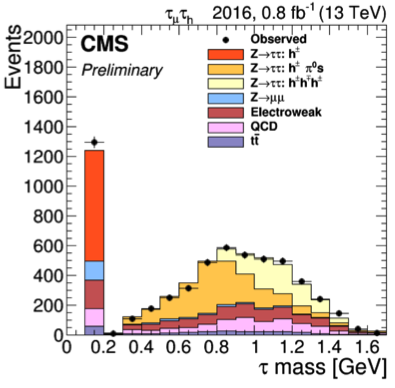}
\caption{ The efficiencies for the generated tau decay modes to be reconstructed as different measured modes (left), the tau mass distributions with Run-1 (middle) and Run-2 (right) data, at CMS \cite{sub2,sub3}. }
\label{fig:fig9}
\end{figure}

Using the $H\to\tau\tau$ decay to measure the CP is experimentally challenging because the neutrinos are not reconstructed. There are two main methods to extract the CP \cite{CP1,CP2,CP3,CP4,CP5}. One is by using the impact parameters to approximately reconstruct the tau decay plane from its leading track. This is best done for the $\tau\to\pi\nu$ decay, and the analyzing power is compromised for the other tau decays. The other is by using the $\tau\to\rho\nu\to\pi^\pm\pi^0\nu$ decay. The tau decay plane can be approximately reconstructed by the track and the neutral pion. However, the relative energy of $\pi^\pm$ and $\pi^0$ need to be classified in order to maximize the analyzing power. In order to use the methods, the tau decay modes (or its substructure) need to be well differenciated. In ATLAS \cite{sub1}, after the initial tau reconstruction and identification, the hadronic tau candidate is further analyzed with a particle flow (PF) algorithm to better reconstruct and differenciate the neutral pions. The energy in the calorimeter deposited by $\pi^\pm$ is then subtracted, and the remaining energy cluster is reconstructed as $\pi^0$ and identified with a Boosted Decision Tree (BDT) method. The final decay mode classification is done with another BDT. The reconstruction and identification efficiencies for a single tau at ATLAS is listed in Tab.~\ref{tab:tab4}. The efficiency for separating the different decay modes and the amount of crosstalk is visualized in Fig.~\ref{fig:fig7}. In general, there is a non-negligible fraction of 2 (1) $\pi^0$ reconstructed as 1 (0) $\pi^0$. In Fig.~\ref{fig:fig8}, the tau energy resolution with the PF is compared with the old baseline method, and the reconstructed tau mass and relative abundance of different modes are compared with data. In conclusion, with the substructure, a factor of 2 improvement of tau energy with respect to the calo-based method at the low $p_T$ (the resolution of the neutral $\pi^0$ is about 16\%), and a factor of 5 improvement in the angular resolution can be achieved. For the neutral $\pi^0$, a precision of 0.006 (0.012) can be reached for it $\eta$ ($\phi$) measurement. 
\begin{table}[h]
\caption{ The Branching Fractions, acceptance, reconstruction and identification efficiencies for a single tau in different decay modes at ATLAS \cite{sub1}. }
\label{tab:tab4}
\centering 
\begin{tabular}{lccc}
\hline
Decay mode & $\mathcal{B}$ [\%] & $\mathcal{A}\cdot\epsilon_{reco}$ [\%] & $\epsilon_{ID}$ [\%] \\ \hline
$h^\pm$ & 11.5 & 32 & 75 \\ 
$h^\pm\pi^0$ & 30.0 & 33 & 55 \\ 
$h^\pm\geq 2\pi^0$ & 10.6 & 43 & 40 \\ 
$3h^\pm$ & 9.5 & 38 & 70 \\ 
$3h^\pm\geq 1\pi^0$ & 5.1 & 38 & 46 \\ \hline
\end{tabular}
\end{table}
CMS also uses the PF constituents (charged and neutral particles) to reconstruct taus \cite{sub2,sub3}. Discriminants are then applied to reject jets and leptons (with a MVA based tau identification). Multiple $\tau_h$ hypotheses for each jet are constructed by a combinatorial approach. The one passing all cuts and with the highest $p_T$ is selected as the decay mode for the tau candidate. The mode reconstruction efficiency, and tau mass with Run-1 and Run-2 data are shown in Fig.~\ref{fig:fig9}. Compared with ATLAS, CMS currently has less decay modes classified, but the crosstalk among different modes is also smaller than ATLAS. Good agreement with Run-1 and Run-2 data is achieved, although the $W$+jets and QCD background are visibly higher in the latter.
Both ATLAS and CMS used the embedding technique to estimate the $Z\to\tau\tau$ background with reduced systematics. This is done by replacing muons from the $Z\to\mu\mu$ data by simulated taus and decaying the taus, and then embedding the simulated tau decay products into the original real data event. No spin correlation information is conserved by this procedure. The Tauola and TauSpinner    packages \cite{tsp} could be used to ensure the correct $Z$ boson polarization and to restore the di-tau spin correlation.

\section{Conclusion}
\label{conclusion}
Testing the CP nature of Higgs is one of the important tasks after its discovery. With the diboson channels ($H\to\gamma\gamma, ZZ, WW$), and assuming pure CP state for the Higgs, the CP even state is favored by data. The tensor structure of the $HVV$ coupling can also be probed for mixed CP scenarios. No significant CP mixing effect is observed and limits are set on the CP-odd terms in the effective Lagrangian. The $H\to\tau\tau$ channel is ideal for probing CP through both the $HVV$ coupling through VBF, and the $H\tau\tau$ Yukawa coupling. Combined sensitivity of ATLAS and CMS already exceeded $5\sigma$ with the Run-1 data. The former has results with VBF $H\to\tau\tau$ based on the Run-1 data. The CP-mixing parameter in the VBF production is currently consistent with zero. The CP mixing in $H\to\tau\tau$ decay can be large in theory, but experimentally very challenging. Both ATLAS and CMS have refined the tau reconstruction with substructure information, which is essential for the CP study in the tau decays. Looking forward, the CP test precision will improve with the HL-LHC data, in which just a few percents uncertainty on the $H\to\tau\tau$ signal strength is expected.

%
%
%

\end{document}